\documentclass[twocolumn,pra,showpacs,floatfix,superscriptaddress]{revtex4-1}%
\usepackage{graphicx}
\usepackage{dcolumn}
\usepackage{bm}
\usepackage{amssymb}
\usepackage{amsmath}
\usepackage{amsfonts}
\usepackage{xcolor}%
\usepackage{float}
\definecolor{darkblue}{rgb}{0.1,0,0.5}
\usepackage[colorlinks,hyperindex]{hyperref}
	\hypersetup
	{
		colorlinks,%
		citecolor=darkblue,%
		linkcolor=darkblue,%
		urlcolor=darkblue,%
	}

\begin{document}

\title{Ionization in Orthogonal Two-Color Laser Fields - Origin and Phase Dependence of Trajectory-Resolved Coulomb Effects}

\author{Martin Richter}
\email{richter@atom.uni-frankfurt.de}
\affiliation{Institut f\"ur Kernphysik, Goethe-Universit\"at Frankfurt, 60438
Frankfurt am Main, Germany}

\author{Maksim Kunitski}
\affiliation{Institut f\"ur Kernphysik, Goethe-Universit\"at Frankfurt, 60438
Frankfurt am Main, Germany}

\author{Markus Sch\"offler}
\affiliation{Institut f\"ur Kernphysik, Goethe-Universit\"at Frankfurt, 60438
Frankfurt am Main, Germany}

\author{Till Jahnke}
\affiliation{Institut f\"ur Kernphysik, Goethe-Universit\"at Frankfurt, 60438
Frankfurt am Main, Germany}

\author{Lothar Ph.H. Schmidt}
\affiliation{Institut f\"ur Kernphysik, Goethe-Universit\"at Frankfurt, 60438
Frankfurt am Main, Germany}

\author{Reinhard D\"orner}
\affiliation{Institut f\"ur Kernphysik, Goethe-Universit\"at Frankfurt, 60438
Frankfurt am Main, Germany}
\date{\today}

\begin{abstract}
We report on electron momentum distributions from single ionization of Ar in strong orthogonally polarized two-color (OTC) laser fields measured with the COLTRIMS technique. We study the effect of Coulomb focusing whose signature is a cusp like feature in the center of the electron momentum spectrum. While the direct electrons show the expected strong dependence on the phase between the two colors, surprisingly the Coulomb focused structure is almost not influenced by the weak second harmonic streaking field. This effect is explained by the use of a CTMC simulation which describes the tunneled electron wave packet in terms of classical trajectories under the influence of the combined Coulomb- and OTC laser field. We find a subtle interplay between the initial momentum of the electron upon tunneling, the ionization phase and the action of the Coulomb field that makes the Coulomb focused part of the momentum spectrum apparently insensitive to the weaker streaking field.
\end{abstract}

\maketitle

\section{Introduction}
In strong field ionization the interaction between the tunneled electron wave packet and the remaining ion has drawn a lot attention in the last two decades. Earlier studies spotlighted the case when the electron wave packet is driven back to the parent ion by the linearly polarized laser field and either collisionally liberates a second electron (Nonsequential Double Ionization, NSDI) \cite{corkum93prl}, recombines and emits a high energy photon (High Harmonic Generation, HHG) \cite{McPherson87josab} or backscatters and is then driven to higher electron energies \cite{Paulus94JPhyB}. During the return of the electron wave packet it is transversally focused by the Coulomb potential towards the ion which results in a higher recollision probability and therefore in a larger rate for NSDI and HHG \cite{Corkum96PRA}. If the total energy of the returning electron is negative it might be trapped into a highly excited Rydberg state \cite{Ivanov01PRA}. More recently the influence of the ionic potential also on the direct, non-returning electrons was investigated by comparing the lateral electron momentum spread in laser fields of different ellipticities \cite{Paulus98prl, Corkum05PRA}. There two kinds of Coulomb focusing were identified. The first one is a weak focusing of the electron wave packet directly after tunneling which exists for all ellipticities and the second one is a focusing accumulated during electron propagation which strongly depends on the polarization of the laser field. In another line of recent studies more complex laser fields like orthogonally polarized two-color pulses (OTC) with a weak second harmonic field were used to steer the electron wave packet dependent on the OTC phase \cite{kitzler05prl,smirnova12nature,richter15prl}.

In this work we bring these two lines of research together and use OTC pulses to study Coulomb focusing and inversely explore the Coulomb focused part of the electron momentum distribution to learn about the dynamics in the joint OTC and Coulomb field. In particular we study how the Coulomb focused part of the electron momentum distribution depends on the phase between the two colors in the OTC field. To this end we investigate single ionization of Ar in a strong OTC field and record the three dimensional electron momentum distribution using a COLTRIMS reaction microscope.

We have found that the cusp like feature in the center of the electron momentum spectrum which is the fingerprint of Coulomb focusing seems to be not influenced by the second harmonic field, in contrast to the rest of the momentum distribution.


In order to shed light on the origin of this observation we use a straightforward model which describes the electron wave packet evolution in the combined laser- and Coulomb-field in terms of trajectories. The effects mentioned before like the influence of the ionic potential on the returning and non-returning electron trajectories and the recollision and recapture mechanisms are described by the model. Our analysis firstly explains details of the trajectories which contribute to the region in momentum space we identify to be resulted from Coulomb focusing. Secondly our analysis shows why this region of phase space dominates over the rest of the momentum distribution.





Before continuing with the more complex OTC-field, the case of linear polarization of the electric field will be discussed.

\begin{figure}[h]
	\centering
	\includegraphics[width=1.0\columnwidth]{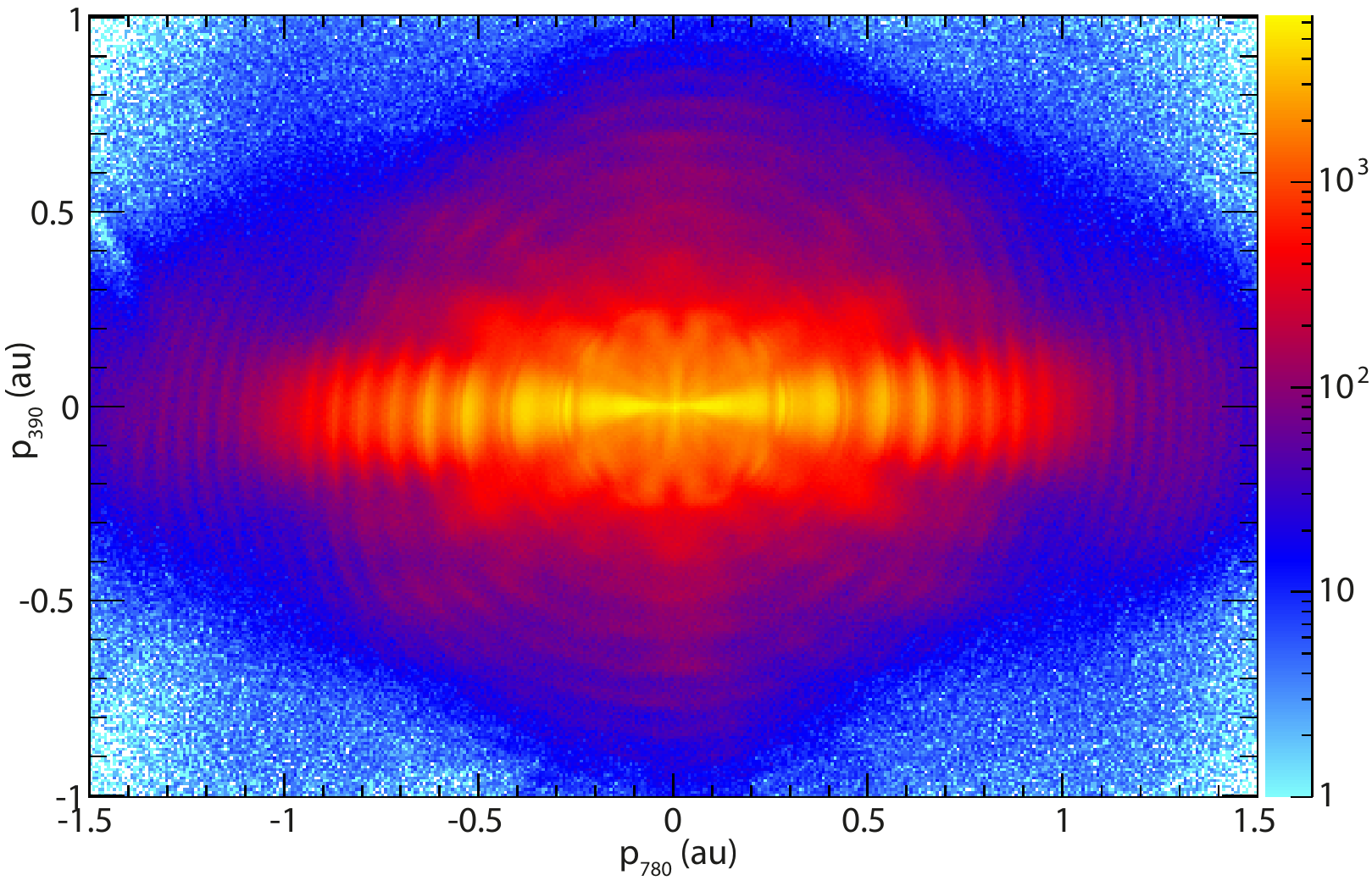}
	\caption{Photoelectron momentum spectrum integrated over $p_{\perp}$ for strong-field single ionization of Ar with a linearly polarized (along $p_{780}$) \mbox{780 nm}, \mbox{40 fs} laser pulse with an intensity of $1.4 \cdot 10^{14} \text{W/cm}^{2}$. Note that for consistency with the following figures the vertical axis is named $p_{390}$ even though there is no \mbox{390 nm} field in this case.}
	\label{fig:introduction}
\end{figure}
Fig. \ref{fig:introduction} shows the electron momentum distribution from single ionization of Ar by a linearly polarized \mbox{780 nm}, \mbox{40 fs} pulse with an intensity of $1.4 \cdot 10^{14} \text{W/cm}^{2}$. The figure contains characteristic structures which are well known from the literature, such as the above-threshold ionization (ATI) structure \cite{agostini79prl} appearing as rings and the \textquotedblleft spider-leg\textquotedblright -shaped holographic interferences \cite{vrakking11science,vrakking12prl,vrakking12prl2}. The electron momentum distribution also shows a cutoff at a momentum of approximately \mbox{$p_z^{cutoff}=1.08$ au} which is the maximum momentum the direct, non-rescattering electrons can acquire in the laser field at this intensity after tunneling. This momentum equals to an energy of $2 U_p$. The higher electron momenta in Fig. \ref{fig:introduction} are understood to originate from the elastic backscattering of the returning electrons forming the \textquotedblleft plateau\textquotedblright \cite{Paulus94JPhyB} and can reach energies up to $10 U_p$. The structure dominating the spectrum is the strong horizontal line along the polarization direction of the laser field. Its narrow transverse momentum distribution is caused by the afore mentioned Coulomb focusing effect which reflects the influence of the left behind ion's potential on the transverse electron momenta. 


\section{Experiment}
\begin{figure}[h]
	\centering
	\includegraphics[width=1.0\columnwidth]{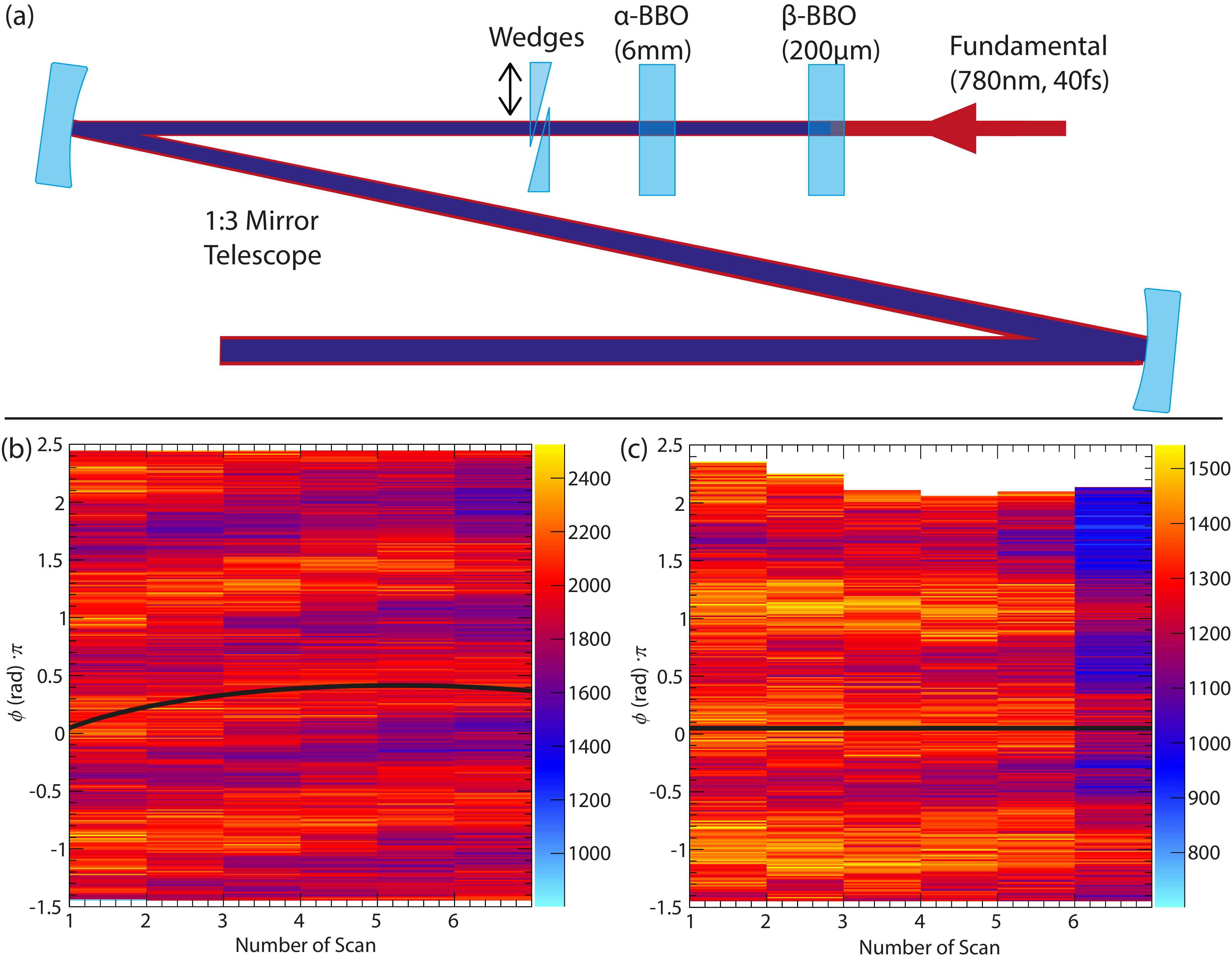}
	\caption{(a) the collinear setup employed for the generation of the OTC-laser field. The second harmonic pulse is obtained in a $\beta$-BBO crystal by frequency doubling the fundamental laser field. Subsequently a $\alpha$-BBO crystal is used for compensation of the group delay between the two-color laser field in the glass of the vacuum chamber viewport and of the wedges. For fine tuning of the phase $\phi$ a pair of fused silica wedges was used. Despite the collinearity a slow ($\pi/2 / 50 \text{min}$) drift of the phase was observed ((b), black curve) and corrected (c) later on during offline analysis.}
	\label{fig:otcsetup_phasekorr}
\end{figure}

The optical collinear setup to generate the OTC field \cite{kitzler05prl,staudtekitzlerprl14,kitzler14pra}
\begin{align} \label{eq-otc-feld}
\vec{E} = E_{780}  \cos(\omega t) \vec{e}_{780} + E_{390}  \cos(2 \omega t + \phi) \vec{e}_{390}
\end{align}
is sketched in Fig. \ref{fig:otcsetup_phasekorr}(a).
 The fundamental \mbox{780 nm} pulses (I=$1.4 \cdot 10^{14} \text{W/cm}^{2}$) were created by a Ti:sapphire multipass amplifier (KMLabs Dragon) at a rate of \mbox{8 kHz}, having a spectral bandwidth of \mbox{28 nm} and a temporal width of \mbox{35 fs} (FWHM). The orthogonally polarized second harmonic field (390 nm, I=$1.3 \cdot 10^{13} \text{W/cm}^{2}$) was generated in a $\beta$-Barium borate-crystal ($\beta$-BBO, \mbox{200 $\mu$m} thick) by frequency doubling of the fundamental \mbox{780 nm} laser pulse which was reduced in diameter by a 3:1 lens telescope to fit the $\beta$-BBO's aperture. To ensure the temporal overlap between the two pulses in this collinear setup an \mbox{6 mm} $\alpha$-BBO crystal (X-cut) and a pair of fused silica wedges were used to compensate for the different group velocities of the OTC pulses. After the wedges both beams were magnified by a 1:3 mirror telescope and steered into the reaction chamber. Because of the dispersion in the beam path the duration of the \mbox{780 nm} and \mbox{390 nm} pulses were stretched to \mbox{55 fs} and to \mbox{70 fs} respectively. 
 
 The phase between the two-color field has been tuned by changing the amount of glass in the beam path. This was achieved by transverally shifting one wedge on the translation stage. Despite the intrinsic stability of a collinear setup we observed slow drifts of the relative phase between the two-color pulses probably resulting from slow changes of temperature and humidity. These drifts were measured and corrected in the offline data-analysis for each ionization event as follows: we continuously scanned the relative phase by shifting the wedge by \mbox{10 $\mu$m} every second. The total travel distance of the stage was \mbox{600 $\mu$m} which covered a phase of around \mbox{4$\pi$}. As soon as the stage reaches the end position it was guided to the starting point for beginning a new scan.The stage position was recorded for each electron measured in the listmode-file. 
 
 The number of counts in a momentum space region ($p_{390}>0.05$au, $|p_{780}|\leq0.1$au) was plotted in a two dimensional histogram with the scan number on the horizontal axis and the apparent phase which was set by the position of the translation stage on the vertical axis. This region in momentum space contains a large amount of streaked electrons, therefore it is very sensitive to the OTC phase. If no drift of the phase occured, this histogram would show horizontal lines, which is however not the case as visible in Fig. \ref{fig:otcsetup_phasekorr}(a).
  
  One scan lasts for approximately 12 minutes which leads to a total time of about 100 minutes along the horizontal axis of the plot in Fig. \ref{fig:otcsetup_phasekorr}(b). Thus, the phase drift during 50 minutes is approximately $\pi/2$. The correction of the phase was done by fitting a line following the streaking maximum indicated by the black curve in Fig. \ref{fig:otcsetup_phasekorr}(b) and subsequent shifting of the maximum for each scan. The result of this correction is shown in Fig. \ref{fig:otcsetup_phasekorr}(c) where the curved maximum distribution has been straightened. 
 
The OTC pulses are focused into a supersonic gas jet of argon atoms and the three dimensional momentum distribution of electrons and ions from single ionization of Argon was measured in coincidence using cold-target recoil-ion momentum spectroscopy (COLTRIMS) \cite{Doerner200095,ullrich2003repprogphys}. The electrons were guided to a time and position sensitive detector \cite{Jagutzki02nuc} by constant electric (\mbox{$10.8$ V/cm}) and magnetic (\mbox{8 Gauss}) fields. With a spectrometer length on the electron side of \mbox{391 mm} electrons emitted within a solid angle of emission of \mbox{4$\pi$} were detected up to a momentum of \mbox{1.5 au}.
 
\section{Simulation}
The calculation is similar to a widely used classical-trajectory Monte Carlo (CTMC) simulation \cite{HuP1997PRA} where the trajectories initial conditions, i.e. the tunneling distance, the initial momentum perpendicular to the electric field vector and the ionization probability were determined by the Ammosov-Delone-Krainov (ADK) theory \cite{Delone91JOSA}. After their launch the simulated electrons perform an oscillatory movement in a six cycle pulse with a trapezoidal envelope consisting of a constant plateau of four cycles and a two cycle ramp down and are propagated $10^8$au in time after the laser pulse is over. Since in \cite{Huang10OptExpress} it is shown that also the long range part of the Coulomb potential is responsible for the focusing effect, the simulated electron trajectories needed to be propagated sufficiently long in time. Ionization occurs only during the first cycle of the pulse.
 \section{Results and Discussion}
\begin{figure}[h]
   	\centering
   	\includegraphics[width=1.0\columnwidth]{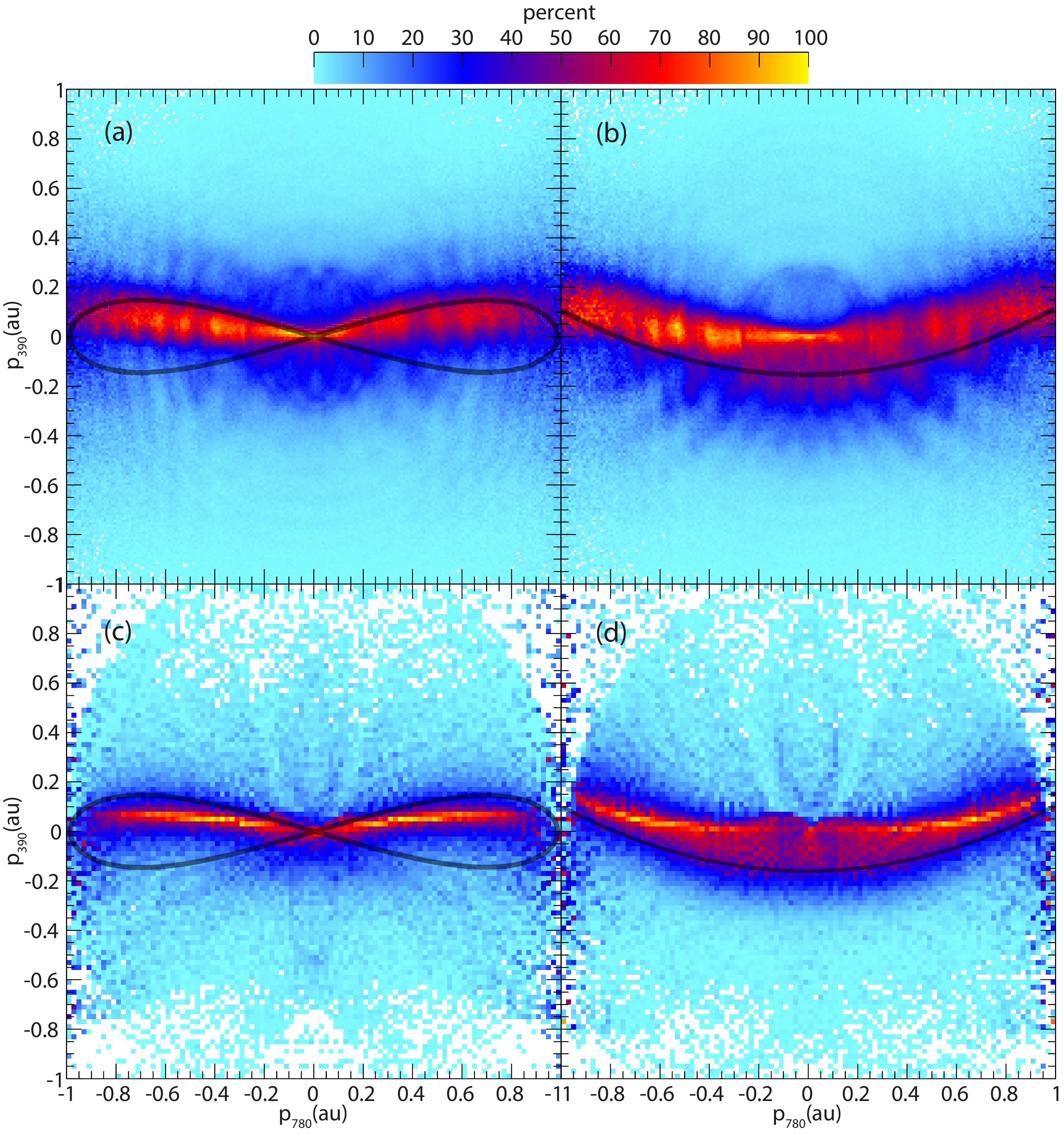}
   	\caption{Measured (a),(b) and calculated (c),(d) photoelectron momentum distribution for single ionization of argon by an OTC-laser field with the intensities $I_{780 nm}=1.4 \cdot 10^{14} \text{W/cm}^{2}$ and $I_{390 nm}=1.3 \cdot 10^{13} \text{W/cm}^{2}$ in polarization plane and the NVP according to the OTC phases. The histograms are normalized column by column so that the sum of the entries in each column equals one. For (a),(c) the phase between the two-color pulses was $\phi=\pi$ and for (b),(d) it was $\phi=3/2 \pi$. The experimental data (a),(b) was recorded as described in the previous chapter and the plots (c),(d) were obtained by a CTMC simulation under consideration of the ionic potential.}
   	\label{fig:2d_exp_theo}
   \end{figure}

 In our experiment the intensity of the ionization field (\mbox{780 nm}) was by a factor of 10 higher than that of the second harmonic field (\mbox{390 nm}). Thus the ionization field governs the tunneling step and dominates the driving of the electrons while the second harmonic field affects the electrons much less. We therefore refer to the second harmonic field as \textquotedblleft streaking field\textquotedblright . The final momentum the electrons gain in the direction of the streaking field after tunneling equals the negative vector potential (NVP) of the streaking field at the instant of ionization. Fig. \ref{fig:2d_exp_theo} shows the measured and calculated electron momenta in polarization plane for the phases \mbox{$\phi=\pi$} (left) and \mbox{$\phi=3/2 \pi$} (right) as well as the according NVP of the two-color field.
 Overlooking the wealth of interference structures, two eye-catching features can be observed. Firstly, a diffuse distribution mainly following the NVP occurs. Electrons contributing to this distribution are called \textquotedblleft direct trajectories\textquotedblright since they only weakly interact with the remaining ionic potential. Secondly, a very prominent line along $p_{390nm}=0$ which is only weakly influenced by the \mbox{390 nm} streaking field can be observed. Especially the distribution in the region around the origin $\{p_{780nm},p_{390nm}\}=\{0,0\}$ does surprisingly not seem to be affected by the \mbox{390 nm} streaking field. 
 
\begin{figure}[h]
 	\centering
 	\includegraphics[width=1.0\columnwidth]{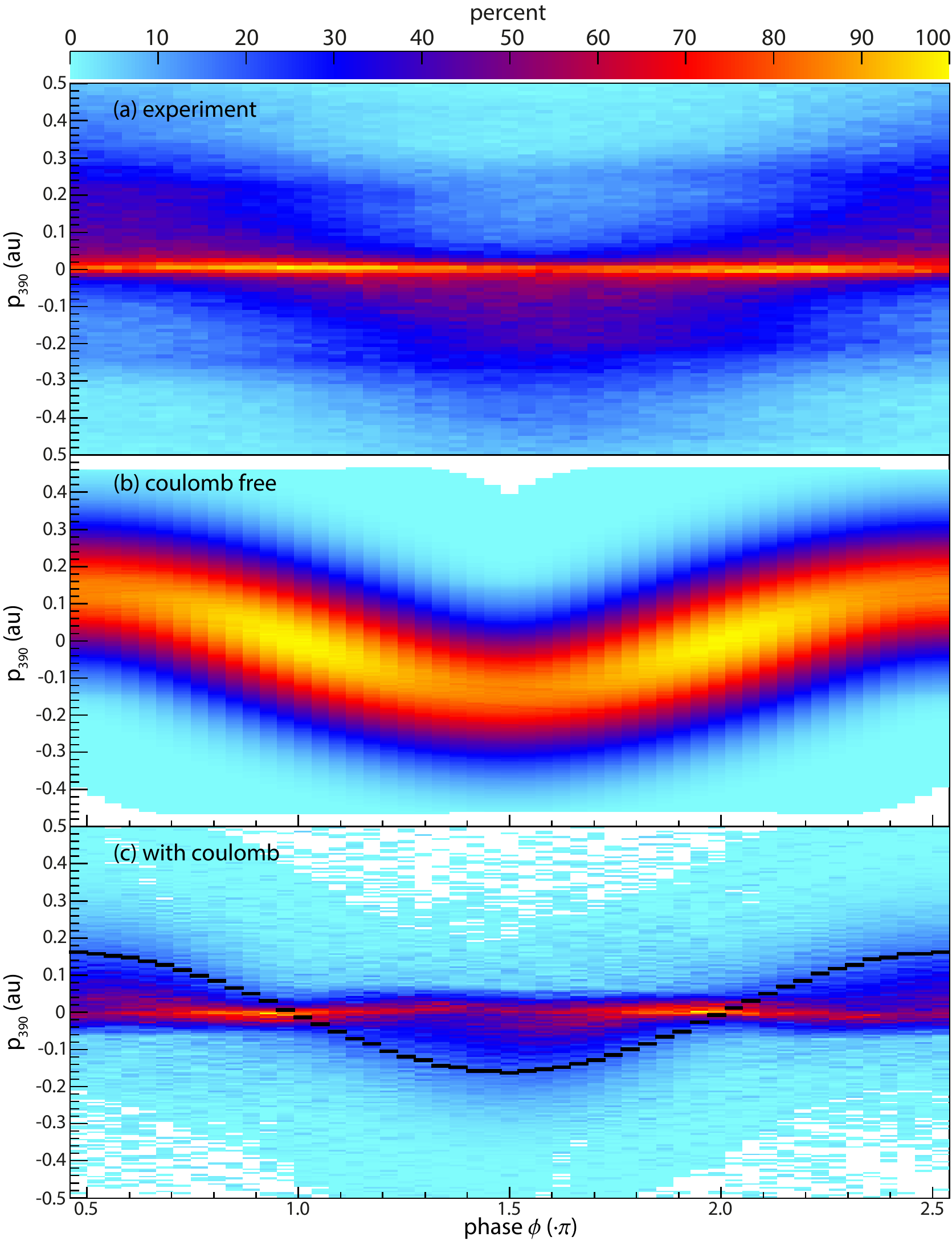}
 	\caption{The dependence of the momentum component along the direction of the \mbox{390 nm} streaking field on the absolute phase in rad between the two-color pulses for the experimental data (a), the Coulomb free simulation (b) and the simulation including the ionic potential (c). In all cases there is a cut in ionization field (\mbox{780 nm}) direction $|p_{780}|\leq0.05$au and also in the direction perpendicular to the polarization plane $|p_{\perp}|\leq0.05$au. In (a) the sinusoidally oscillating blue halo and the unstreakable narrow red line can be observed. Additionally in (c) the NVP of electrons which start at the maximum of the ionization field is shown by a black line. As expected it has the same trend as the curve in (b).}
 	\label{fig:phase_py}
 \end{figure}

Both features can also be seen in Fig. \ref{fig:phase_py}(a) where the phase $\phi$ versus the \mbox{390 nm} momentum component $p_{390}$ for a small gate $|p_{780}|\leq0.05$ au is plotted. The direct trajectories are represented by the sinusoidally oscillating blue halo. In the analysis they were used for the calibration of the absolute phase between the two-color pulses. In contrast to the direct trajectories in Fig. \ref{fig:phase_py}(a) there is a straight red line along $p_{390nm}=0$ which does not oscillate. Independently of the phase the electrons forming this prominent feature seem not to be affected by the \mbox{390 nm} streaking field.

 To learn more about the origin of the \textquotedblleft unstreakable\textquotedblright part of the spectrum in Fig. \ref{fig:phase_py}(a) we performed a simulation which describes the propagation of the electron trajectories after strong field ionization under the influence of the two-color laser field and an optional ionic potential in three dimensions. The sinusoidally oscillating blue halo is already reproduced by the Coulomb free simulation Fig. \ref{fig:phase_py}(b) since it originates from trajectories which leave the ions vicinity without strong interaction with the ionic core and therefore following the NVP. Only when including the Coulomb potential to the calculation Fig. \ref{fig:phase_py}(c) both features, namely the \textquotedblleft unstreakable\textquotedblright red line and the blue halo are reproduced.
 
\begin{figure}[h]
  	\centering
  	\includegraphics[width=1.0\columnwidth]{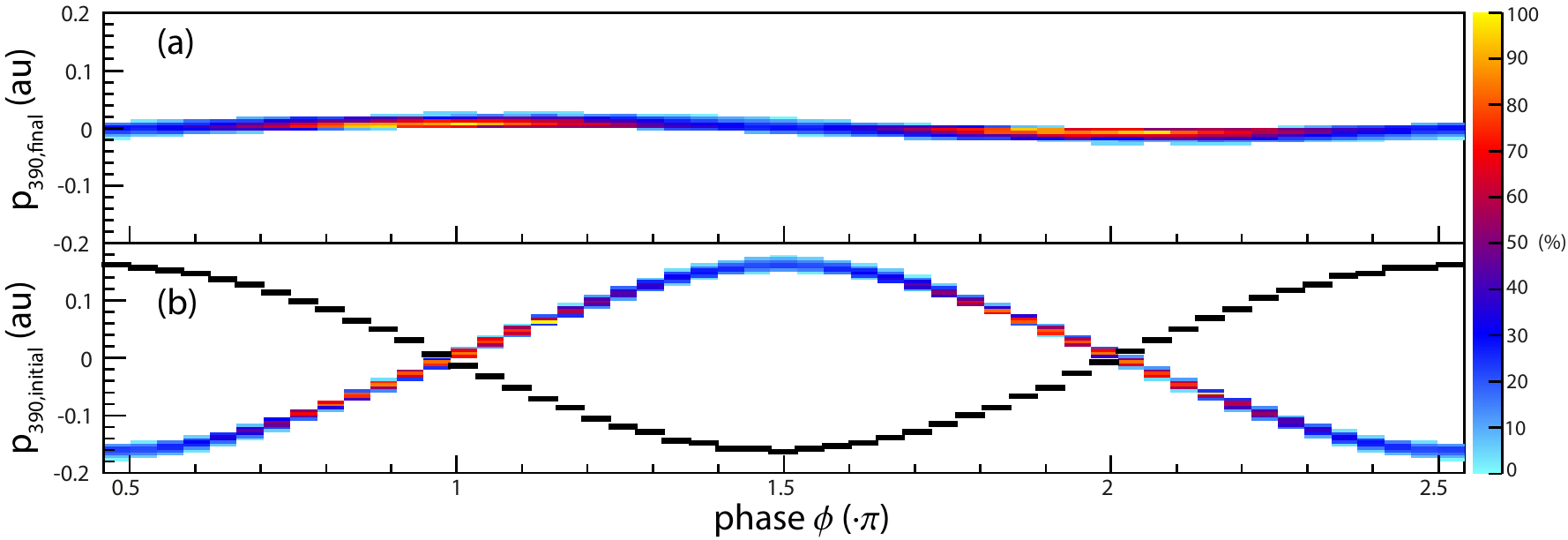}
  	\caption{Simulated results for the Coulomb field free case where the final momentum (a) or the initial momentum (b) in the streaking field direction versus the absolute phase between the two-color laser field is plotted. For both graphs only those trajectories were selected which passed the core within a radius of \mbox{1.5 au}. The resulting final momentum dependence is a narrow line which corresponds to the measured central line of Fig. \ref{fig:phase_py}(a). The initial momentum dependence (b) is exactly opposite to the NVP of the \mbox{390 nm} streaking field (black line). This implies that the momentum gained by electrons in the streaking field cancels completely the initial momentum upon tunneling in the same direction.}
  	\label{fig:phase_vs_py_nocoulomb}
  \end{figure}

 Firstly, the \textquotedblleft Coulomb free case\textquotedblright (i.e. the case neglecting the Coulomb potential) will be discussed in order to get a better understanding of the streaking dynamics. The broad curve in Fig. \ref{fig:phase_py}(b) simply arises due to the NVP of the two-color laser field at the instant of ionization and the width of the initial momentum distribution. As expected only the measured sinusoidal blue halo formed by the electrons which only weakly interact with the remaining core is reproduced by the Coulomb free simulation. The \textquotedblleft unstreakable\textquotedblright part of the spectrum is expected to result from electrons which strongly interact with the ion because of coming close to it at some time during the laser cycles. In the simulation, by tracking the path of the electrons in the laser field, their minimum distance from the ion can be determined. Consequently if in the simulation only those trajectories which came close to the core ($\approx$ \mbox{1.5 au}) are taken into account, a set of trajectories which would strongly interact with the ion can be selected approximatively. The final momentum component along the \mbox{390 nm} field for this set of trajectories as a function of the phase between the streaking and ionization field is shown in Fig. \ref{fig:phase_vs_py_nocoulomb}(a). As in Fig. \ref{fig:phase_py} we have selected only events with $|p_{780}|\leq0.05$au and $|p_{\perp}|\leq0.05$au.
  
  Surprisingly, without further conditions this set of trajectories nicely resembles the secondly mentioned \textquotedblleft unstreakable\textquotedblright central feature of the experimental data. Examination of the initial momentum distribution (at the tunnel exit) of these trajectories (Fig. \ref{fig:phase_vs_py_nocoulomb}(b)) reveals that this set of trajectories has the initial momentum which is exactly opposite to the NVP of the streaking field (black curve). This means that for electrons, which pass close by the nucleus under the condition of no Coulomb field, the momentum gained by the \mbox{390 nm} streaking field is just compensated by their initial momentum in this direction.
  
This compensation mechanism holds however for the Coulomb free case only. Including the Coulomb potential into the calculations makes the scenario much more complex. The simulated phase dependence of the electron momentum under consideration of the ionic potential is depicted in Fig. \ref{fig:phase_py}(c) where the two features, namely the sinusoidal blue halo formed by the direct trajectories and the \textquotedblleft unstreakable\textquotedblright red line agree well to those of the experiment. Additionally the NVP of the electrons born at the ionization field maximum is sketched by a black curve in the figure. The blue halo which relates to the direct electrons clearly follows this curve. 

\begin{figure}[h]
  	\centering
  	\includegraphics[width=1.0\columnwidth]{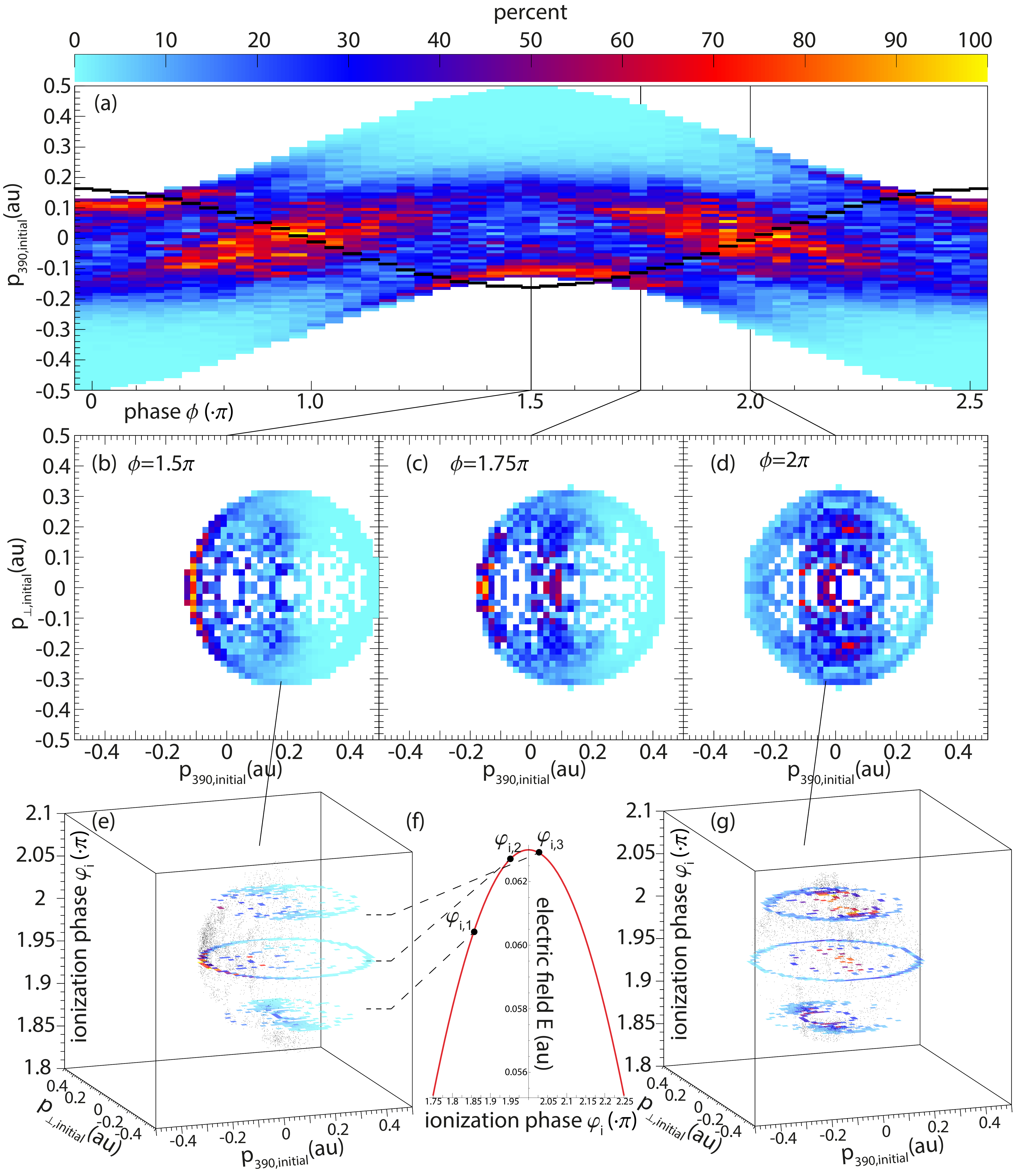}
  	\caption{Initial conditions of trajectories leading to the cusp structure in the momentum space. The final electron momentum was restricted to the range $\{|p_{390}|,|p_{780}|,|p_{\perp}|\}\leq\{0.01,0.05,0.05\}$. (a) The initial momentum in \mbox{390 nm} streaking field direction versus the OTC phase. (b),(c),(d) initial electron momenta in the plane perpendicular to the \mbox{780 nm} ionization field for the OTC phases as indicated in the figure. The 3D distributions in (e),(g) were constructed by plotting the dependence of the initial momenta of (b) and (d) on the ionization phase $\varphi_i$ with respect to the ionization field. The ionization phases $\varphi_i$ illustrated in (f) correspond to the instances of ionization in the \mbox{780 nm} field.	
  	The black line in the panel (a) represents the NVP of the \mbox{390 nm} streaking field. With the help of the initial condition map in (e) and (g) three types of trajectories could be identified. The direct ones born between $\varphi_{i,1}$ and $\varphi_{i,2}$ and the indirect ones launched between $\varphi_{i,2}$ and $\varphi_{i,3}$ start with momenta located on the spherical shells (e),(g). The third type of trajectories are referred to as the intermediate electrons which originate between $\varphi_{i,1}$ and $\varphi_{i,3}$ with initial momenta located within the sphere.}
  	\label{fig:phase_pinit}
  \end{figure}

To get an insight into the formation of the cusp like distribution along $p_{390nm}=0$ the initial momentum distribution of the corresponding trajectories in the presence of the Coulomb field is examined. Therefore in Fig. \ref{fig:phase_pinit}(a) the initial momenta $p_{390nm,init}$ in the streaking field direction with a condition on the final momenta \mbox{$|p_{390}|\leq0.01$ au}, \mbox{$|p_{780}|\leq0.05$ au}, \mbox{$|p_{\perp}|\leq0.05$ au} versus the phase are shown. Here $p_{\perp}$ denotes the momentum component perpendicular to the polarization plane, i.e. out of the plane shown in Fig. \ref{fig:2d_exp_theo}.
The phase dependence of the initial momenta in the streaking field direction (Fig. \ref{fig:phase_pinit}(a)) does not resemble that of the Coulomb field free case where the streaking momentum due to the \mbox{390 nm} electric field (black curve in Fig. \ref{fig:phase_pinit}(a)) compensates the initial momentum distribution. However the inclusion of the Coulomb potential leads to a large width of initial momenta of about \mbox{$\Delta p_{390,init} \approx 0.6$ au} of electrons that contribute to the cusp like structure. Somehow during their propagation in the combined laser- and Coulomb field these electrons are focused to a narrow final momentum width of \mbox{$\Delta p_{390,final}=0.04$ au}. Another effect of the Coulomb potential is that the mean value of the initial distribution does slightly follow the NVP of the streaking field instead of being opposite to it which means that the Coulomb potential overcompensates the initial momenta.



To get a better picture of the Coulomb interactions we will consider now the two dimensional initial momentum distributions for three selected OTC phases (Figs. \ref{fig:phase_pinit}(b)-(d)). Examining the development of the initial momentum distribution for increasing phases in Fig. \ref{fig:phase_pinit}(b)-(d) a notable shift of the whole distribution can be observed. For the phase $\phi=3/2 \pi$ the circle is centered at $p_{\perp,init}=0$ and p$_{390,init}$=0.18au and when increasing the phase the whole distribution shifts to p$_{390,init}$=0. This shift represents the afore mentioned compensation mechanism of the Coulomb free case since the shift almost exactly compensates the \mbox{390 nm} electric fields streaking momentum (black curve in \ref{fig:phase_pinit}(a)). However the situation is completely different if the color coded yield of the starting momenta is considered. 

\begin{figure}[h]
  	\centering
  	\includegraphics[width=1.0\columnwidth]{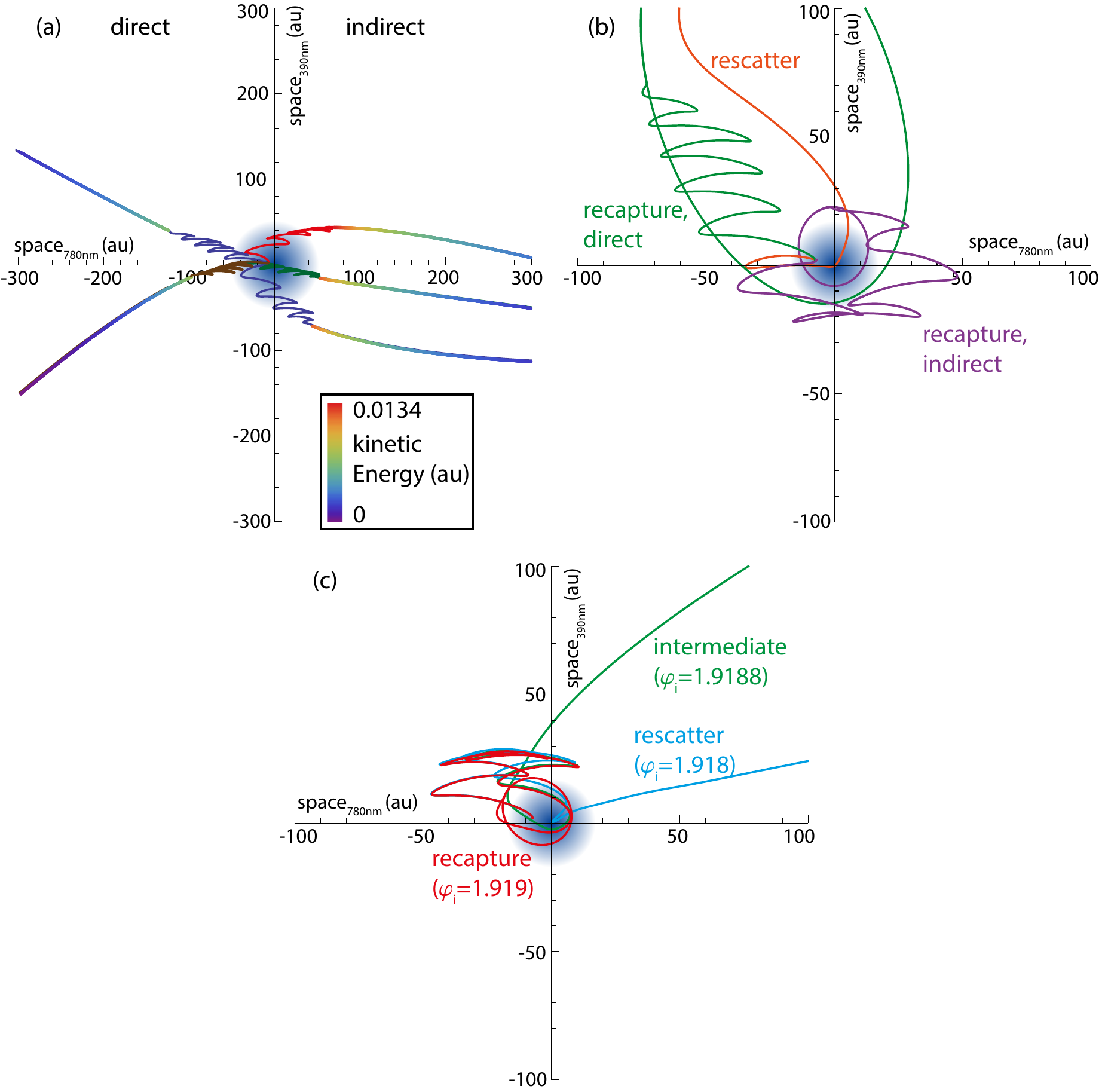}
  	\caption{Evolution of different groups of electron trajectories in polarization plane. (a) typical \textquotedblleft direct\textquotedblright and \textquotedblleft indirect\textquotedblright trajectories leading to zero final momentum. After the laser pulse these electrons are decelerated by the Coulomb attraction of the ionic core which results in the loose of kinetic energy (see the color code). The time the trajectories need for the displayed paths in (a) is in the order of \mbox{$10^3$ au} while in the full calculation they are propagated \mbox{$10^8$ au} to ensure that the remaining potential energy is negligible. Panel (b) shows trajectories corresponding to the recapture and rescatter mechanisms which define the inner radius of the spherical shells of Fig. \ref{fig:phase_pinit}(e),(g). Note that also the direct electrons can be recaptured if their kinetic energy at the end of the laser pulse is too small. (c) typical intermediate trajectory (green) and the extremely different outcomes when changing the ionization phase $\varphi_i$ only minusculy.
  		}
  	\label{fig:Pfade_direct_indirect}
  \end{figure}

To be able to understand this behavior the paths of the electrons need to be identified and therefore the tunneling phases $\varphi_i$ need to be brought into play. For the OTC phases $\phi=3/2 \pi$ and $\phi=2 \pi$ the dependence of the initial momentum distribution in the plane perpendicular to the ionizing \mbox{780 nm} field on the ionization phase is depicted in Fig. \ref{fig:phase_pinit}(e) and \ref{fig:phase_pinit}(g) respectively. These three-dimensional (3D) histograms are like a map in the \textquotedblleft momentum-tunneling time\textquotedblright space of the relevant initial conditions of the Coulomb focused trajectories. We have identified three types of such trajectories. The two main types are located on a spherical shell in the \textquotedblleft momentum-tunneling time\textquotedblright space and are called the \textquotedblleft direct-\textquotedblright and \textquotedblleft indirect electrons\textquotedblright which are well known from the literature. The former are born within the ionization phases $\varphi_{i,1}$ and $\varphi_{i,2}$ of the laser pulse (\ref{fig:phase_pinit}(f)) and leave the vicinity of the ion without returning to it. Typical evolutions of direct trajectories in coordinate space leading to final zero momentum are depicted on the left of \ref{fig:Pfade_direct_indirect}(a). The indirect electrons (on the right of \ref{fig:Pfade_direct_indirect}(a)) are liberated between the ionization phases $\varphi_{i,2}$ and $\varphi_{i,3}$ of the laser pulse (\ref{fig:phase_pinit}(f)) and pass once nearby the nucleus. The third type of electrons start with momenta which are not located on the spherical shell but lie rather inside of the sphere. During propagation they undergo complex movements (\ref{fig:Pfade_direct_indirect}(c),green) and are entitled as \textquotedblleft intermediate electrons\textquotedblright in the following.

A very delicate interplay of initial conditions is needed for the direct and indirect electrons to reach zero final momentum. The boundaries of the spherical shell are given by the following two scenarios. The first one is when the electrons are driven away by the laser field from the ionic potential too fast and therefore gain high final momenta which defines the outer radius of the shell. In the second case the electrons fail to gain enough energy to escape from the ion after the laser pulse is over leading to recapture into a highly excited Rydberg state (Fig. \ref{fig:Pfade_direct_indirect}(b,purple and green). The indirect electrons additionally have a chance to scatter off the ion which would lead to high final momenta (Fig. \ref{fig:Pfade_direct_indirect}(b,orange). The recapture or rescatter mechanisms determine the radius of the inner boundary since these mechanisms correspond to the large extent of the initial conditions inside the sphere.

The initial conditions for the intermediate electrons form only small islands within the spheres of Figs. \ref{fig:phase_pinit}(e,g). In no case they will gain enough momentum after tunneling to escape the ions vicinity like the direct and indirect electrons without making detours. But it is still possible for them just to escape the ion loosing almost the whole energy that was gained in the field. An example for one of these intermediate trajectories is the green one shown in Fig. \ref{fig:Pfade_direct_indirect}(c) with the initial conditions $\{p_{\perp,init};p_{390,init};\varphi_i\}=\{0;0.15;1.9188 \pi \}$. A tiny deviation from this small island of initial conditions will produce two completely different outcomes. An increase of the ionization phase to $\varphi_i=1.919 \pi$ leads to the recapture of an electron \ref{fig:Pfade_direct_indirect}(c,red), while in contrast a decrease of the ionization phase to $\varphi_i=1.918 \pi$ results in a trajectory which scatters to a high final momentum of about \mbox{1 au} \ref{fig:Pfade_direct_indirect}(c,blue).

\begin{figure}[h]
  	\centering
  	\includegraphics[width=1.0\columnwidth]{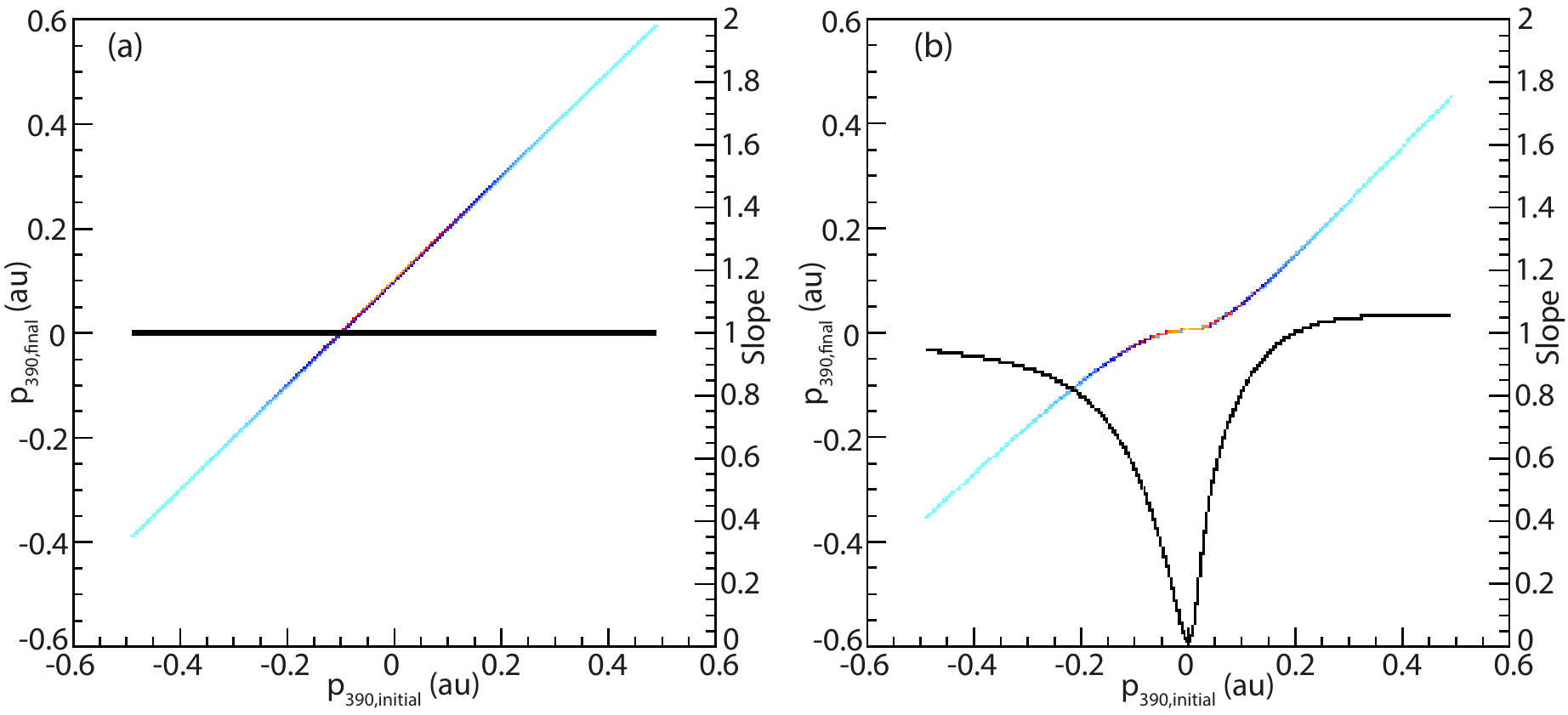}
  	\caption{The initial versus the final momentum in \mbox{390 nm} streaking field direction with the OTC phase $\phi=1.95 \pi$, an ionization phase of $\varphi_i=1.87 \pi$, an initial momentum perpendicular to the polarization plane of \mbox{$p_{\perp,init}=0.06$ au} and a Gaussian momentum distribution along $p_{390,initial}$ ranging from -0.5 to \mbox{0.5 au}. In the Coulomb free case (a) the correlation between initial and final momentum is linear (colored line) resulting in a constant slope for this line along $p_{390}=0$ which axis is located on the right side of the plot. The black line is the derivative of the deflection function. The final momentum distribution is shifted to larger momenta according to the NVP of the \mbox{390 nm} streaking field. For the same initial conditions but under consideration of the Coulomb potential (b) the colored curve around $p_{390}=0$ is very flat. Again the slope of the colored curve is represented by the black line which yields almost zero for the flat region. The color encodes the electron density.}
  	\label{fig:Abbildungsfunktion}
  \end{figure}

Up to this point the initial conditions of the trajectories which end in the cusp like feature at $p_{390nm}=0$ have been identified. This does not yet explain why there is a cusp, i.e. an enhancement of the photoelectron density in this Coulomb focused region. As the distribution of initial conditions set by the tunneling process is a Gaussian, i.e. a smoothly varying function, cusps in the final momentum distribution results from a zero in the derivative of the deflection function mapping the initial phase to the final momentum space. We illustrate this for an exemplary set of trajectories (Fig. \ref{fig:Abbildungsfunktion}) which are ionized at the phase $\varphi_{i}=1.87 \pi$ with the streaking phase $\phi=1.95 \pi$ starting with a Gaussian momentum distribution in streaking field direction and a fixed initial momentum in the direction perpendicular to the polarization plane \mbox{$p_{\perp,init}=0.06$ au}. For the Coulomb free case in Fig. \ref{fig:Abbildungsfunktion}(a) the initial versus the final momentum in streaking field direction is represented as a diagonal colored line. As expected the initial momentum distribution is shifted to higher momenta by the NVP of the streaking field. Since for a fixed ionization time all trajectories are shifted by the same amount by the streaking field, the slope of the colored line equals one (black horizontal line, right scale). 
When switching on the Coulomb potential for the same set of trajectories the resulting correlation between initial and final momentum in streaking field direction is shown in Fig. \ref{fig:Abbildungsfunktion}(b). The corresponding colored curve is very flat in the region around $p_{390}=0$ which is also reflected by the black slope curve which drops almost to zero at this point. In this case a broad range of initial momenta \mbox{$|p_{390,initial}|\leq0.05$ au} lead to almost the same momenta around $p_{390}=0$. This focusing affects the trajectories from all ionization phases and leads, compared to the Coulomb free case, to a net increase of the electron contribution to zero final momentum. Note that the OTC phase for Fig. \ref{fig:Abbildungsfunktion} is chosen arbitrarily and very similar figures could be made for all other streaking phases.

\section{Conclusions}
Throughout the paper we have studied the Coulomb focusing of electrons upon ionization in the orthogonally polarized two-color laser field. Firstly, we have found that the focused part whose signature is the cusp like structure in the photoelectron momentum spectrum is not streaked by the \mbox{390 nm} electric field. However the cusp is composed of trajectories with initial momentum distributions which strongly depend on the OTC phase. Secondly, we explained the domination of the focused part over the rest in the final momentum electron distribution. This was done by means of a classical simulation in which the propagation of the photoelectrons after ionization was described in terms of classical trajectories. This work is a further step of understanding the ionization and electron propagation in OTC laser fields which could help to gain better control of processes like HHG and Laser Induced Electron Diffraction.

We acknowledge support by the Deutsche Forschungsgemeinschaft in the frame of the Priority Programme \textquotedblleft Quantum Dynamics in Tailored Intense Fields\textquotedblright.



\bibliographystyle{apsrev4-1}
\bibliography{complete}






\end{document}